\documentstyle[twocolumn,aps,epsfig,floats]{revtex}

\draft 

\begin{document} 
\twocolumn[\hsize\textwidth\columnwidth\hsize\csname @twocolumnfalse\endcsname
\title{Unquenched large orbital magnetic moment in NiO} 
\author{ S. K. Kwon and B. I. Min } 
\address{ Department of Physics, 
          Pohang University of Science and Technology, 
          Pohang 790-784, Korea } 
\date{February 22, 2000}
 
\maketitle 
 
\begin{abstract}    
Magnetic properties of NiO are investigated by incorporating the
spin-orbit interaction in the LSDA + $U$ scheme.
It is found that the large part of orbital moment remains unquenched in NiO. 
The orbital moment contributes about $\mu_L = 0.29\mu_B$ 
to the total magnetic moment of $M = 1.93 \mu_B$,
as leads to the orbital-to-spin angular momentum ratio of $L/S = 0.36$. 
The theoretical values are in good agreement 
with recent magnetic X-ray scattering measurements.
\end{abstract}

\pacs{PACS number: 71.20.Be, 71.70.Ej, 75.10.Lp, 75.50.Ee}
]

Electronic and magnetic structures of late $3d$ transition metal (TM) 
mono-oxides (MnO, FeO, CoO, and NiO) have been extensively investigated
over the last decades \cite{Mattheiss}.
Conventional band theory with the local spin density approximation 
(LSDA) fails to describe the electronic structures 
of the compounds.  In the LSDA calculations,
the energy gaps of MnO and NiO are underestimated\cite{Terakura}.
Even worse, FeO and CoO are predicted to be metallic
which are, on the contrary, large gap insulators in nature.
The main problem in the LSDA is the use of mean field-type 
exchange-correlation functional 
which is improper to describe localized $3d$ electrons.

There have been several theoretical efforts 
to cure the deficiencies in the LSDA, for example, 
the self-interaction correction (SIC) scheme \cite{Cowan,Zunger,Svane}, 
the GW approximation (GWA) \cite{Hedin,Aryasetiawan},
and the LSDA + $U$ method \cite{Anisimov,Liechtenstein}.
The SIC-LSDA is in the line of the extended LSDA 
by removing unphysical electron interaction with itself.
In the GWA, the quasi-particle energy is obtained 
through the self-energy calculation to the lowest order in
the screened Coulomb interaction $W$.
The GWA method applied to NiO gives a rather good description 
of the energy gap size \cite{Aryasetiawan}.
Computational load, however, is very heavy in the GWA.
The LSDA + $U$ method overcomes 
the failure of the LSDA by incorporating the on-site Coulomb correlation $U$ 
of the multiband Hubbard model-like.
In this method, localized $3d$ electrons are treated 
separately from delocalized $sp$ electrons.
As a result, all the $3d$ TM mono-oxides in the LSDA + $U$ are 
obtained as insulators with well developed energy gaps which are comparable 
to experimental values \cite{Anisimov}.
In this way, the energy gap problem in $3d$  TM mono-oxides 
is considered to be solved by various calculational methods. 

Due to the outermost characteristics of  TM $3d$ electrons,
atomic $3d$ orbitals are greatly deformed in solids
by the crystal field and/or band hybridization effects.
Hence, the orbital moment of $3d$ TM ion is
usually quenched in solids, because it originates from atomic nature of 
involved atomic elements\cite{Kittel}.
For example, all the $3d$ ferromagnetic transition metals 
of Fe, Co, and Ni show negligible
orbital magnetic moments in the range of $\mu_L \lesssim 0.1 \mu_B$ \cite{Min}. 
It is, however, expected that the strong Coulomb correlation 
in $3d$ TM mono-oxides preserves
the orbital moment of localized $3d$ electrons 
by reducing the ligand crystal field effects at metal ion sites.

In fact, for CoO, it is easily conceived 
that the orbital moment is only partially quenched 
because the measured magnetic moment $M = 3.4 \mu_B$ \cite{Khan} 
simply exceeds the spin magnetic moment alone and 
the minority $t_{2g}$ band is occupied only two-thirds of its available states.
Therefore, the existence of orbital moment in CoO has been stressed
many times \cite{Terakura,Svane,Shishidou}.  
It is also shown that the LSDA + $U$ method gives 
a good description of magnetic structure of CoO 
with a large orbital moment of $\mu_L \sim 1 \mu_B$ \cite{Solovyev}.
On the other hand, in the case of NiO, 
measured magnetic moments are in the range 
of $M = 1.77 \sim 2.2\pm0.2 \mu_B$ \cite{Fender,Cheetham,Fernandez}, 
which are comparable to the spin only moment of isolated Ni$^{2+}$ ion. 
Therefore, the magnetic moment in NiO has been fully attributed 
to the spin moment and the orbital moment is 
expected to be completely quenched.

However, recent magnetic X-ray scattering measurement \cite{Fernandez}
indicates that the orbital-to-spin angular momentum ratio in NiO is 
as large as $L/S = 0.34$, far from fully quenched orbital moment.
In the nonresonant magnetic X-scattering,
the separation of spin and orbital moment is possible, 
because the spin and orbital moment densities have different geometrical 
prefactors in the scattering cross section that can be adjusted 
by changing either the scattering geometry or the X-ray polarization.
In Ref. \cite{Fernandez}, the orbital moment in NiO is extracted 
by the polarization analysis of nonresonant magnetic-scattering intensities.  
This method has evidenced a large contribution of the orbital moment 
to the total magnetic moment.
They also found that the spin and orbital moments in NiO are collinear.

Magnetic moment is one of the basic ground state quantities 
which should be provided by an appropriate band method. 
If the orbital moment is so large in NiO, 
all the previous attempts \cite{Anisimov,Towler,Shick} 
which tried to directly compare calculated spin moments 
with experimental moments are under a mistake. 
The linearized muffin-tin orbital (LMTO) calculation for NiO 
with an orbital-polarization (OP) correction (LSDA + OP) 
in a crystal-field basis yields 
the spin and orbital magnetic moments of $\mu_S = 1.43 \mu_B$ 
and $\mu_L = 0.12 \mu_B$, respectively \cite{Norman1}.
Although the total magnetic moment is significantly improved 
to $M = 1.55 \mu_B$ with the OP correction, 
as compared to $M = 1.23 \mu_B$ from the LSDA,
it is still smaller than the experimental measurements 
and the ratio of $L/S = 0.17$ is only half of 
the magnetic X-ray scattering value \cite{Fernandez}.  
The SIC-LSDA method \cite{Svane} yields 
a large orbital moment of $\mu_L = 0.27\mu_B$,
which, however, was suspected as a methodological artifact 
by other authors \cite{Anisimov,Towler}.
In more recent study, the orbital moment is simply neglected \cite{Shick}. 

\begin{figure}[t] 
\epsfig{file=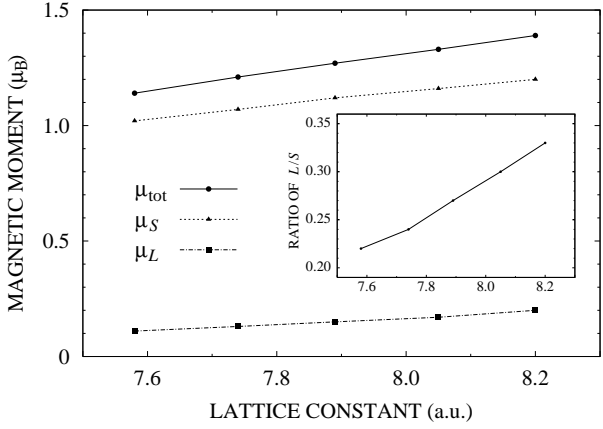,width=8.65cm}
\caption{\label{fig1} 
Magnetic moment behavior with varying the
lattice constant in the LSDA. Inset is the behavior 
of the orbital-to-spin angular momentum ratio $L/S$. 
Increase of all the magnetic moments is understood 
by the reduction of the crystal field and the band hybridization strengths
with volume expansion. 
}
\end{figure}
\begin{figure}[t] 
\epsfig{file=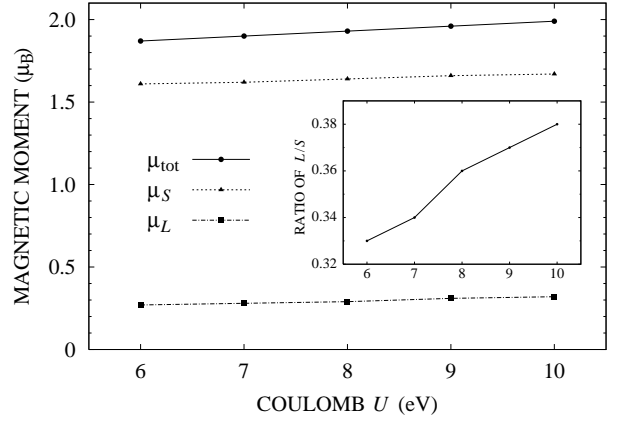,width=8.65cm}
\caption{\label{fig2} 
Coulomb correlation effects on the magnetic moments obtained 
by the LSDA + $U$ method.  
We have used the exchange parameter of $J = 0.89$ eV.
Inset is the orbital-to-spin angular momentum ratio $L/S$. 
The total magnetic moment $M = 1.93 \mu_B$ and the ratio $L/S = 0.36$ 
for $U = 8.0$ eV are in good agreement with experimental values.
}
\end{figure}
To determine the size of orbital moment in NiO,  
we have performed the LSDA + $U$ calculations 
using the LMTO band method within the atomic sphere approximation (ASA).
The incorporation of the spin-orbit coupling into the LSDA + $U$
method is known to give right orbital polarization 
in strongly correlated electron systems \cite{Solovyev}.
The LSDA + $U$ Hamiltonian is given by 
\begin{equation}
{\cal H}_{{\mathrm LSDA} + U} = {\cal H}_{\mathrm LSDA}
                              - {\cal H}_{\mathrm dc} + {\cal H}_{U}
\end{equation}
where the first term in the right hand side is the LSDA Hamiltonian 
and the second term is the double counting correction 
for the third term ${\cal H}_{U}$.
With the Coulomb interaction $U$ and exchange interaction $J$ parameters,
one can write ${\cal H}_{\mathrm dc}$ and ${\cal H}_{U}$, respectively, as 
\begin{eqnarray}
{\cal H}_{\mathrm dc} &=& \frac{1}{2}UN(N-1)  
    - \frac{1}{2}J\sum_{\sigma} N^{\sigma}(N^{\sigma}-1), \\ 
{\cal H}_{U} &=& \frac{1}{2}\sum_{\{m\},\sigma}V(mm';m''m''')
                             n_{mm''}^{\sigma}n_{m'm'''}^{-\sigma} \nonumber \\ 
    &+& \frac{1}{2}\sum_{\{m\},\sigma}
        \left [ V(mm';m''m''')-V(mm';m'''m'') \right ] \nonumber \\
& &~~~~~~~~~\times n_{mm''}^{\sigma} n_{m'm'''}^{\sigma}
\end{eqnarray}
where $n_{mm'}^{\sigma}$ is the $d$ occupation number matrix 
of spin $\sigma$ and $N^{\sigma} = {\mathrm Tr}(n_{mm'}^{\sigma})$,
$N = N^+ + N^-$.
We relate the screened Coulomb interaction $V(mm';m''m''')$
with  the Slater integral $F^k$;
\begin{eqnarray}
V(mm';m''m''') = \sum_{k=0}^{2l}c^{k}(lm,lm'')c^{k}(lm',lm''')F^{k},
\end{eqnarray}
where $c^k(lm,lm')$ is a Gaunt coefficient.
For $3d$ electrons, three Slater integrals of $F^0, F^2$, and $F^4$
are involved in the calculation. 
Among those, the ratio of $F^4/F^2$ is known to be constant 
$\sim 0.625$ for most $3d$ TM atoms \cite{Sawatzky}. 
Hence, the actual number of parameters is 
reduced from three $F^k$'s to two of $U$ and $J$, 
which are given by $U = F^0$ and $J = (F^2 + F^4)/14$, respectively.
To determine the orbital moment, the spin-orbit coupling is simultaneously 
included in the self-consistent variational loop \cite{Min}.
We have assumed the antiferromagnetic ordering state of type-II, 
and the experimental lattice constant of $a=7.893$ a.u. is used.

\begin{table*}[t]
\caption{\label{table1}
Calculated spin ($\mu_S$), orbital ($\mu_L$), and total ($M$) magnetic 
moments in $\mu_B$ at the experimental lattice constant of $a=7.893$ a.u..
$L/S$ is the orbital-to-spin angular momentum ratio. }
\begin{tabular}{cccccc}
Method   &LSDA$^a$&LSDA + OP$^b$&SIC-LSDA$^c$&LSDA + $U^d$&Experiment\\\hline 
$\mu_S$  & 1.12 & 1.43        & 1.53         & 1.64  & 1.90$\pm 0.2^e~$ \\
$\mu_L$  & 0.15 & 0.12        & 0.27         & 0.29  & 0.32$\pm 0.05^e$ \\
$M$      & 1.27 & 1.55        & 1.80         & 1.93  & 1.77$^f$, 1.90$^g$,
                                                       2.2$\pm$0.2$^e$ \\
$L/S$    & 0.27 & 0.17        & 0.35         & 0.36  & 0.34$^e$
\end{tabular}
$^a$ Present results. \\  
$^b$ Reference \cite{Norman1}. \\ 
$^c$ Reference \cite{Svane}. \\
$^d$ Present results with $U = 8.0$ eV and $J = 0.89$ eV. \\
$^e$ Reference \cite{Fernandez}. \\
$^f$ Reference \cite{Fender}. \\
$^g$ Reference \cite{Cheetham}. 
\end{table*}
In Fig.~\ref{fig1}, magnetic moment behavior is shown 
with varying the lattice constant in the LSDA. 
Both the spin and orbital magnetic moments increase monotonically
with increasing the lattice constant. 
This feature is understandable in view of that the crystal field strength 
at Ni sites and the hybridization between Ni $3d$ and O $2p$ bands
are reduced with volume expansion.
At the experimental lattice constant of $a=7.893$ a.u., 
the spin and orbital magnetic moments are obtained 
as $\mu_S = 1.12 \mu_B$ and $\mu_L = 0.15 \mu_B$, respectively, 
which are consistent with existing results \cite{Norman1}.
Both the spin and orbital magnetic moments are larger in NiO than in fcc Ni 
which has $\mu_S = 0.63 \mu_B$ and $\mu_L = 0.06 \mu_B$.
This suggests that $3d$ electrons are more localized in NiO 
than in Ni metal.
Although the ratio of $L/S = 0.27$ in the LSDA is only slightly smaller 
than $L/S = 0.34$ in the experiment (see Table.~\ref{table1}), 
the spin and orbital polarizations  in the LSDA
are not large enough because of underestimation 
of the Coulomb correlation between $3d$ electrons. 

Figure~\ref{fig2} shows the Coulomb correlation effects 
on the magnetic moments obtained by the LSDA + $U$ method. 
We have used the exchange parameter of $J = 0.89$ eV 
which is comparable to literature value \cite{Anisimov}.
Once the strong Coulomb interaction is introduced between Ni $3d$ electrons, 
the magnetic moments increase substantially.
The role of the Coulomb interaction is significant 
to localize Ni $3d$ electrons.
With $U = 8.0$ eV found in Ref. \cite{Anisimov} and Ref. \cite{Norman2}, 
we have obtained the spin and orbital magnetic
moments of $\mu_S = 1.64 \mu_B$ and $\mu_L = 0.29 \mu_B$, respectively.
The total magnetic moment $M = 1.93 \mu_B$ 
and the ratio $L/S = 0.36$ for $U = 8.0$ eV are 
in good agreement with the experimental data.
The change of the Coulomb parameter $U$ by $1.0$ eV 
results in increments of both the magnetic moments 
and the ratio $L/S$ by $\Delta\mu_S \sim \Delta\mu_L \sim 0.01 \mu_B$ 
and $\Delta (L/S) \sim 0.01$, respectively.
The overall total magnetic moment difference 
and the ratio $L/S$ difference are 
only $\Delta M = 0.12 \mu_B$ and $\Delta (L/S) = 0.05$ 
inbetween $U = 6.0$ and $10$ eV.
Thus, all the magnetic moments and the ratio $L/S$ are rather insensitive
to the Coulomb interaction $U$ within the employed parameter range. 

Calculated magnetic moments are summarized in Table.~\ref{table1}
in comparison with experimental data and previous results 
by other calculational methods. 
As mentioned above, 
the sizes of magnetic moments are underestimated in the LSDA.
The total magnetic moment of $M = 1.27 \mu_B$ in the LSDA is 
much smaller than $M = 1.77 \sim 2.2\pm0.2 \mu_B$ 
in experiments \cite{Fender,Cheetham,Fernandez}.
In the LSDA + OP, the spin moment is significantly 
improved to $\mu_S = 1.43 \mu_B$ \cite{Norman1},  
which, however, is still smaller than the experimental values.
More serious is the decrease of orbital moment in the LSDA + OP,
which unfavorably makes the consistency between the theoretical 
and experimental $L/S$ ratio become worse.
Both the SIC-LSDA ($M = 1.80 \mu_B$) \cite{Svane} 
and the LSDA + $U$ ($M = 1.93 \mu_B$) 
give values in agreement with experimental data.
But it can be concluded that the LSDA + $U$ results with
$U = 8.0$ eV and $J = 0.89$ eV, 
among various calculations, are especially 
in best agreement with experimental measurements. 
The ratio of $L/S = 0.36$ in the LSDA + $U$ is also consistent with
$L/S = 0.34$ in the magnetic X-ray scattering measurement \cite{Fernandez}. 

In conclusion, we have found using the LSDA + $U$ calculations 
that the orbital moment in NiO is not fully quenched, which is
as large as $\mu_L = 0.29 \mu_B$.
Both the total magnetic moment of $M = 1.93 \mu_B$ 
and the orbital-to-spin angular momentum
ratio of $L/S = 0.36$ for $U = 8.0$ eV are close to experimental values.
The Coulomb correlation and the spin-orbit coupling are
crucial to get right magnetic polarization in NiO.

Acknowledgements$-$
The authors would like to thank K.B. Lee for helpful discussions. 
This work was supported by the KOSEF (1999-2-114-002-5)
and in part by the Korean MOST-FORT fund.

 

\begin{thebibliography}{99} 
\bibitem{Mattheiss} L. F. Mattheiss, Phys. Rev. B {\bf 5}, 290 (1972).
\bibitem{Terakura} K. Terakura, A. R. Williams, T. Oguchi,
                   and J. K$\ddot{\mathrm u}$bler,
                   Phys. Rev. Lett. {\bf 52}, 1830 (1984);
                   K. Terakura, T. Oguchi, A. R. Williams, 
                   and J. K$\ddot{\mathrm u}$bler,
                   Phys. Rev. B {\bf 30}, 4734 (1984).
\bibitem{Cowan} R. D. Cowan, Phys. Rev. {\bf 164}, 54 (1967);
                I. Lindgren, Int. J. Quantum Chem. {\bf 5}, 411 (1971).
\bibitem{Zunger} A. Zunger, J. P. Perdew, and G. L. Oliver, 
                 Solid State Commun. {\bf 34}, 933 (1980); 
                 J. P. Perdew and A. Zunger, 
                 Phys. Rev. B {\bf 23}, 5048 (1981).
\bibitem{Svane} A. Svane and O. Gunnarsson, 
                Phys. Rev. Lett. {\bf 65}, 1148 (1990).
\bibitem{Hedin} L. Hedin and S. Lundqvist, in {\it Solid State Physics},
                edited by H. Ehrenreich, F. Seitz, and D. Turnbell
                (Academic, New York, 1969), Vol 23, p. 1.
\bibitem{Aryasetiawan} F. Aryasetiawan and O. Gunnarsson,
                       Phys. Rev. Lett. {\bf 74}, 3221 (1995).
\bibitem{Anisimov} V. I. Anisimov, J. Zaanen, and O. K. Andersen,
                   Phys. Rev. B {\bf 44}, 943 (1991).  
\bibitem{Liechtenstein} A. I. Liechtenstein, V. I. Anisimov, and J. Zaanen,
                        Phys. Rev. B {\bf 52}, R5467 (1995).
                        For review, see V. I. Anisimov, F. Aryasetiawan, 
                        and A. I. Liechtenstein,
                        J. Phys. : Condens. Matter {\bf 9}, 767 (1997).
\bibitem{Kittel} C. Kittel, {\it Introduction to Solid State Physics} 7th ed.,
                 (John Wiley \& Sons, New York, 1996), Chapter 14.
\bibitem{Min} B. I. Min and Y.-R. Jang,  
              J. Phys.: Condens. Matter {\bf 3}, 5131 (1991).
\bibitem{Khan} D. C. Khan and R. A. Ericksson, 
               Phys. Rev. B {\bf 1}, 2243 (1970).
\bibitem{Shishidou} T. Shishidou and T. Jo, J. Phys. Soc. Jpn.
	{\bf 67}, 2637 (1998).
\bibitem{Solovyev} I. V. Solovyev, A. I. Liechtenstein, and K. Terakura,
                   Phys. Rev. Lett. {\bf 80}, 5758 (1998).
\bibitem{Fender} B. E. F. Fender, A. J. Jacobson, and F. A. Wegwood, 
                 J. Chem. Phys. {\bf 48}, 990 (1968).  
\bibitem{Cheetham} A. K. Cheetham and D. A. O. Hope, 
                   Phys. Rev. B {\bf 27}, 6964 (1983).  
\bibitem{Fernandez} V. Fernandez, C. Vettier, F. de Bergevin, C. Giles, 
                    and W. Neubeck, Phys. Rev. B {\bf 57}, 7870 (1998); 
                    W. Neubeck, C. Vettier, V. Fernandez, F. de Bergevin, 
         and C Giles, J. Appl. Phys. {\bf 85}, 4847 (1999). 
\bibitem{Towler} M. D. Towler, N. L. Allan, N. M. Harrison, V. R. Saunders,
                 W. C. Mackrodt, and E. Apr$\grave{\mathrm a}$,
                 Phys. Rev. B {\bf 50}, 5041 (1994).
\bibitem{Shick} A. B. Shick, A. I. Liechtenstein, and W. E. Pickett,
                Phys. Rev. B {\bf 60}, 10 763 (1999).
\bibitem{Norman1} M. R. Norman, 
                  Phys. Rev. B {\bf 44}, 1364 (1991). 
\bibitem{Sawatzky} F. M .F. de Groot, J. C. Fuggle, B. T. Thole, and
                   G. A. Sawatzky, Phys. Rev. B {\bf 42}, 5459 (1990).
\bibitem{Norman2} M. R. Norman and A. J. Freeman,
                  Phys. Rev. B {\bf 33}, 8896 (1986).
\end{thebibliography}
\end{document}